# Effect of sigma electrons on the pi-electron behavior in single-wall carbon nanotubes

G.-P. Tong* and Q.-P. Huang

*Department of Physics, Zhejiang Normal University, Jinhua 321004, Zhejiang, China*

The hybrid orbitals of single-wall carbon nanotubes are given according to the structure of the nanotube. Because the energy levels of these hybrid orbitals are close to each other, the σ orbitals will affect the behavior of the π electrons, which is called the scattering of π electrons. This scattering effect is taken into account in the nanotube and the local wave function of π electrons is constructed, which is called the extended Wannier function. In the Wannier representation, the electronic hopping energies and the energy gap of the tubes (9,0) and (9,9) are calculated. Our results show that the band gap of the tubes increases in direct ratio with the scattering coefficients of σ orbitals and this scattering is able to enhance the localization of π electrons.

As a simple one-dimensional nanoscaled structure, single-wall carbon nanotubes (SWNTs) have been the subject of intensive theoretical and experimental efforts because of their singular properties and potential applications. In the past years, the studies about the electronic properties of SWNTs have had rapid developments. The energy band structure has been studied by the first-principles techniques using the local-density approximation[1,2] and the transfer matrix method [3]. The density of states in the vicinity of the Fermi level of SWNTs can be expressed in terms of a universal relationship that depends only on whether the nanotube is metallic or semiconducting[4]. The electronic densities of states of atomically resolved SWNTs have been investigated using scanning tunneling microscopy[5]. Hopping energies plays an important role in the studying about electronic structure of SWNTs. Many researches about the electronic structure of SWNTs have been carried on under the frames of SSH (Su-Schrieffer-Heeger) model. However, the usual SSH model doesn't take the π-σ hybridization effect into account. Because the energy of σ orbitals is very close to the π-orbital energy, the behavior of π-electrons will be subject to the scattering of σ orbitals. Therefore, the scattering influences on the energy gap of a SWNT. When we study the electronic structure of SWNTs, the same hopping integral $\gamma_0$ has usually been defined [6]. In fact, it isn't proper to define the same hopping energy $\gamma_0$. The different hopping integrals $\gamma_i$ take the curvature effect into account in Ref[3]. However, the energy structure is different from those by other approaches when the hopping energies are defined like that.

In this paper we calculate the hopping energy and the energy gap of SWNTs in the Wannier representation. However, it is very difficulty to gain the rigorous Wannier function. We will use the Kohn's method [7] to construct the Wannier function. Because the lattice-atom wave function is not orthonormal, we can obtain the orthogonal and normalized basic function by combining the lattice-atom wave functions. This function has a good feature of localization and it can be considered as the Wannier function, called the extended Wannier function. When we construct the Wannier fuction, the effect of σ electrons on the π electrons is included, and which is called the scattering of σ electrons. The hopping energies and band gaps of the tubes (9,0) and (9,9) are calculated. Our results show that the band gaps of the tubes increase in direct ratio with the scattering coefficients of σ orbitals and this scattering is able to enhance the locality of the π electrons.

It is well-known that graphite follows the theory of $sp^2$ and π electrons only include the pure $2p_z$ orbital. However, for a SWNT the orbital is no longer normal to the surface of the tube and has a small slanting angle deviation [8]. According to symmetry, the hybridization orbitals have one π-orbital and three σ-orbitals. The π-electron can hop between the lattice atoms. The lattice atom wave functions are not orthogonal to each other. That is to say, the atom wave function can't be considered as the Wannier function. Therefore, if the single lattice atom function is defined as $\varphi$, the local wave function of the $i$th lattice can be written as[9]:

$$|W_i\rangle = b_{ii}|\varphi_i\rangle + \sum_m b_{im}|\varphi_m\rangle \qquad (m \neq i) \qquad (1)$$

where the summation of $m$ is over all sites except for the $i$th lattice. The coefficient $b_{im}$ can be obtained from following equation:

$$\delta_{ij} = \langle W_i | W_j \rangle$$
$$= b_{ii}b_{jj}S_{ij} + \sum_n b_{ii}b_{jn}S_{in} + \sum_m b_{jj}b_{im}S_{jm} + \sum_{nm} b_{jn}b_{im}S_{nm} \qquad (2)$$

where $S_{ij} = \langle \varphi_i | \varphi_j \rangle$, means the overlap integral between the $i$th and $j$th lattices.

According to the orthogonality and completeness of the Bloch function, the Wannier functions are composed of the orthogonal and perfect functions. But the Wannier functions between the different lattices are orthogonal. However, different bands have different Wannier functions. The Wannier functions not only have the localization features but also perfect features. Equation (1) can be approximately considered as the Wannier function for the same energy band. The Hamiltonian of π-electrons can be written as:

$$H = \sum_{i,s} \varepsilon_{i,s} C^+_{i,s} C_{i,s} + \sum_{ij,s} \gamma_{ij}(C^+_{j,s}C_{i,s} + H.C.) \qquad (3)$$

where $\varepsilon_i$ represents the orbit energy of π electrons, $\gamma_{ij}$ is the hopping energy between the $i$th lattice and the $j$th lattice. In the Wannier representation, we define $\varepsilon_i$ and $\gamma_{ij}$ as follows:

$$\varepsilon_i = \langle W_i | H | W_i \rangle \qquad (4)$$

$$\gamma_{ij} = \langle W_i | H | W_j \rangle \qquad (5)$$

where $H$ is given by

$$H = -\frac{1}{2}\nabla^2 - \sum_n \frac{Z_n}{r_n} \qquad (6)$$

Here $H$ represents the single-electron Hamiltonian in atom units. $Z_n$ is the effective nuclear charge number. We should emphasize that the Bloch function is the eigenfunction of Hamiltonian for the perfect crystals and the Wannier function is not the eigenfunction.

It is common knowledge that the π electron only includes the pure $2p_z$ orbit in graphite. The orbits will be rehybridized when the graphite is rolled into a SWNT. The π-electrons maybe have a probability to take up $2s$ orbits. It is proper to describe the orbits of a SWNT using the unequal $sp^3$ hybridization. The wave functions of π electrons are composed of $2s$ and $2p_z$ orbits. The four hybridized orbitals can be expressed as:

$$|\sigma_i\rangle = \sqrt{A_i}|s\rangle + \sqrt{1-A_i}\left(a_{i1}|p_x\rangle + a_{i2}|p_y\rangle + a_{i3}|p_z\rangle\right), \quad (i=1,2,3)$$

$$|\pi\rangle = \sqrt{A_4}|s\rangle + \sqrt{1-A_4}|p_z\rangle \quad (7)$$

and

$$A_1 + A_2 + A_3 + A_4 = 1$$

where $A_i$ denotes the $s$-orbital component in three σ-orbitals, $A_4$ is the $s$-orbital component in the π-orbitals, and $a_{ij}$ are the coefficients of the $p$-orbital components. On the basis of the geometry of a SWNT, the unequal $sp^3$ hybridization orbitals can be obtained:

$$\begin{pmatrix}|\sigma_1\rangle \\ |\sigma_2\rangle \\ |\sigma_3\rangle \\ |\pi\rangle\end{pmatrix} = \begin{pmatrix} \sqrt{A_1} & \sqrt{2/3} & 0 & -\sqrt{(1-3A_1)/3} \\ \sqrt{A_1} & -1/\sqrt{6} & \sqrt{2}/2 & -\sqrt{(1-3A_1)/3} \\ \sqrt{A_1} & -1/\sqrt{6} & -\sqrt{2}/2 & -\sqrt{(1-3A_1)/3} \\ \sqrt{1-3A_1} & 0 & 0 & \sqrt{3A_1} \end{pmatrix}\begin{pmatrix}|s\rangle \\ |p_x\rangle \\ |p_y\rangle \\ |p_z\rangle\end{pmatrix} \quad (8)$$

where $|s\rangle, |p\rangle$ are the hydrogen-like atom wave functions, which are given by

$$|s\rangle = \sqrt{\frac{\lambda^3}{\pi}}(1-\lambda r)e^{-\lambda r}, \quad |p_x\rangle = \sqrt{\frac{\lambda^5}{\pi}}xe^{-\lambda r}, \quad |p_y\rangle = \sqrt{\frac{\lambda^5}{\pi}}ye^{-\lambda r}, \quad |p_z\rangle = \sqrt{\frac{\lambda^5}{\pi}}ze^{-\lambda r} \quad (9)$$

Here $\lambda$ is the Slater orbit index number and reflects the shield conditions of the inner layer electrons. Because the $s$-orbital component increases with decreasing of the radius of a SWNT, the Slater orbit index number can be approximately expressed as

$$\lambda \approx \lambda_0\left(1+\sqrt{1-3A_1}\right) \approx \lambda_0\left(1+\frac{\sqrt{2}a_{c-c}}{4r}\right) \quad (10)$$

where $\lambda_0$ is the Slater orbit index of graphene, $a_{c-c}$ is the carbon-carbon bond length, and $r$ denotes the radius of a SWNT. In the light of quantum theory, π electrons will be subject to the scattering of other orbits and have a probability to take up these σ-orbitals. In this sense, the π-electron wave function can be expressed by these hybridized orbitals:

$$|\varphi\rangle = k_1|\sigma_1\rangle + k_2|\sigma_2\rangle + k_3|\sigma_3\rangle + k_4|\pi\rangle, \quad (11)$$

and

$$k_1^2 + k_2^2 + k_3^2 + k_4^2 = 1$$

where $k_i$ (i=1,2,3) are called the scattering coefficients of the σ-orbitals. Substituting (11) into (1) can obtain the Wannier function required.

To show the effect of the σ-orbitals, we assume that the scattering coefficients $k_i$ are equal for three σ-orbitals, i.e. $k_i = k$, in which (11) is reduced to the form as follows

$$|\varphi\rangle = \left(3k\sqrt{A_1} + \sqrt{(1-3k^2)(1-3A_1)}\right)|s\rangle + \left(\sqrt{3A_1(1-3k^2)} - k\sqrt{3(1-3A_1)}\right)|p_z\rangle \quad (12)$$

Fig.1 shows a part of the zigzag carbon nanotubes $(n, 0)$. In our calculation, we take the bond lengths[10] $a_{c-c}= 0.1408$ nm and the orbit index number $\lambda_0=1.82$. For the tube (9,0), we can obtain $\lambda=2.08$ from (10). When $k$ is equal to 0.1, three nearest hopping energies and the corresponding energy gaps are given in Table I as a function of the effective nuclear charge number $Z$. From Table I we can see that the magnitude of the hopping energies decreases and the energy gap increases with the increase of $Z$. This shows that when the influence of σ orbitals is taken into account the increasing of $Z$ results in the electron-localization weakening. On the other hand, the difference value of $|\gamma_3|-|\gamma_2|$ increases as $Z$. Their competition leads to the increasing of the gap. When $Z$ is fixed, the hopping energies and the gap varying as the coefficient $k$ are listed in Table II. From Table II, for the same value $Z$ the magnitude of the hopping energies and the energy gap increase with $k$ increasing, and what's more, the values of $|\gamma_3|-|\gamma_2|$ also increase with $k$. This means that the σ-orbitals are able to widen the gap. This is because the π-electron is subject to the scattering of the σ orbitals, that is to say, there is a probability occupying the σ orbitals for the π-electrons, which can enhance the localization of the π-electron and then the energy gap becomes large. From the above we conclude that hybrid orbitals have important influence on the hopping energies and the energy gap of a SWNT. In order to show this further the tube (9,9) is calculated. From (10), we have $\lambda=1.97$. When $k = 0.1$, the nearest hopping energies and the energy gap of the tube (9, 9) are listed in Table III. Table IV shows the relationship between the hopping energies and the scattering coefficient $k$ and between the gaps and the coefficient $k$. By comparison, we see that this effect of the tube (9,9) is similar to that of the tube (9,0), i.e. the σ orbitals can change the localization of the π-electrons in SWNTs.

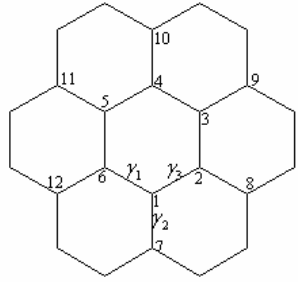

FIG.1. A part of the structural scheme for the zigzag carbon nanotubes $(n, 0)$, where $\gamma_1$, $\gamma_2$, and $\gamma_3$ denote the hopping energies of π-electrons.

TABLE I. $k = 0.1$, the nearest hopping energies and the energy gaps of the tube (9, 0) change with the effective nuclear charge number $Z$.

| $Z$ | $\gamma_1 = \gamma_3$ (eV) | $\gamma_2$ (eV) | $|\gamma_3|-|\gamma_2|$ | Energy gap (eV) |
| --- | --- | --- | --- | --- |
| 2.870 | -2.703 | -2.622 | 0.081 | 0.162 |
| 2.880 | -2.685 | -2.602 | 0.083 | 0.166 |
| 2.890 | -2.667 | -2.582 | 0.085 | 0.170 |
| 2.900 | -2.649 | -2.562 | 0.087 | 0.174 |
| 2.910 | -2.631 | -2.542 | 0.088 | 0.178 |

TABLE II. $Z$=2.890, the nearest hopping energies and the energy gaps of the tube (9, 0) change with the scattering coefficient $k$.

| $k$ | $\gamma_1 = \gamma_3$ (eV) | $\gamma_2$ (eV) | $\lvert\gamma_3\rvert - \lvert\gamma_2\rvert$ | Energy gap (eV) |
| --- | --- | --- | --- | --- |
| 0.000 | -2.433 | -2.383 | 0.050 | 0.100 |
| 0.100 | -2.667 | -2.582 | 0.085 | 0.170 |
| 0.135 | -2.782 | -2.678 | 0.104 | 0.208 |

TABLE III. $k$= 0.1, the nearest hopping energies and the energy gaps of the tube (9, 9) change with the effective nuclear charge number $Z$.

| $Z$ | $\gamma_1$ (eV) | $\gamma_2 = \gamma_3$ (eV) | $\lvert\gamma_1\rvert - \lvert\gamma_2\rvert$ | Energy gap (eV) |
| --- | --- | --- | --- | --- |
| 2.870 | -2.578 | -2.335 | 0.243 | 0.486 |
| 2.880 | -2.561 | -2.316 | 0.245 | 0.490 |
| 2.890 | -2.545 | -2.297 | 0.248 | 0.496 |
| 2.900 | -2.528 | -2.278 | 0.250 | 0.500 |
| 2.910 | -2.511 | -2.259 | 0.252 | 0.504 |

TABLE IV. $Z$=2.890, the nearest hopping energies and the energy gaps of the tube (9, 9) change with the scattering coefficient $k$.

| $k$ | $\gamma_1$ (eV) | $\gamma_2 = \gamma_3$ (eV) | $\lvert\gamma_1\rvert - \lvert\gamma_2\rvert$ | Energy gap (eV) |
| --- | --- | --- | --- | --- |
| 0.000 | -2.365 | -2.143 | 0.222 | 0.444 |
| 0.100 | -2.545 | -2.297 | 0.248 | 0.496 |
| 0.135 | -2.646 | -2.383 | 0.263 | 0.526 |

The nearest neighbor hopping energy $\gamma_0$ is usually from -2.5eV to -2.7eV in the experiments[11]. In our work, when $Z$=2.890, $\lambda$=2.08, and $k_1 = k_2 = k_3 = 0.1$, the energy gap is 0.171eV and very close to the result by using the first-principle method[1]. Our method has two good points, a clear physical pattern and a little calculating time, compared to the first-principle method. In addition, the Slater orbit indices of SWNTs of different radii can be given and the smaller the radius, the larger the index and the better the localization of the π-electrons. Therefore, we think that the Wannier function method is a good method in the calculation of the electronic properties for SWNTs.

## Acknowledgement

This work is supported by the Natural Science Foundation of Zhejiang Province,China(Grant No Y605167)

*tgp6463@zjnu.cn

[1] T. Miyake, S. Saito, *Phys. Rev.* B **72,** 073404(2005)

[2] V. Zólyomi and J. Kürti, *Phys. Rev.* B **70,** 085403(2004)

[3] J. W. Ding, X. H. Yan, and J. X. Cao, *Phys. Rev.* B **66,** 073401(2002)

[4] J. W. Mintmire and C. T. White, *Phys. Rev. Lett.* **81,** 2506(1998)

[5] P. Kim, T. W. Odom, J. L. Huang, and C. M. Lieber, *Phys. Rev. Lett*. **82,** 1225(1999)

[6] S. Reich, J. Maultzsch, C. Thomsen, and P. Ordejón, *Phys. Rev.* B **66,** 035412(2002)

[7] W. Kohn, *Phys. Rev.* B **7,** 4388(1973)

[8] A. Kleiner and S. Eggert, *Phys. Rev.*B **64,** 113402(2001)

[9] G. P. Tong, *Chinese Journal of Semiconductors* **20,** 365(1999)( in Chinese)

[10] J. Kurti, G. Kresse, and H. Kuzmany, *Phys. Rev.* B **58,** R8869(1998)

[11] M. Ouyang, J. L. Huang, and C. M. Lieber, *Phys. Rev. Lett.* **88,** 066804(2002)

[12] R. Saito, M. Fujita, G. Dresselhaus, and M. S. Dresselhaus, *Appl. Phys. Lett.* **60**, 2204(1992)